\documentstyle[12pt]{article}
\def\v1{\vspace{1cm}}
\def\be{\begin{equation}}
\def\ee{\end{equation}}
\def\bc{\begin{center}}
\def\ec{\end{center}}
\def\ik{\partial}
\def\vh{\varphi}

\newcommand{\bea}{\begin{eqnarray}}
\newcommand{\eea}{\end{eqnarray}}

\begin{document}
\title
{Bogoliubov Quasiparticles in Constrained Systems}

\author{ V.N.Pervushin, V.I.Smirichinski \\[0.3cm]
{\normalsize\it Joint Institute for Nuclear Research},\\
 {\normalsize\it 141980, Dubna, Russia.} }

\date{\empty}
\maketitle
\medskip

\begin{abstract}
{\large
The paper is devoted to the formulation of quantum field theory for
an early universe in General Relativity considered as
the Dirac general constrained system.
The main idea is the Hamiltonian reduction of the constrained system
in terms of measurable
quantities of the observational cosmology: the world proper time,
cosmic scale factor, and the density of matter.
We define " particles" as field variables in the holomorphic
representation which diagonalize the measurable density.
The Bogoliubov quasiparticles are determined by  diagonalization of the
equations of motion (but not only of the initial Hamiltonian) to get
the set of integrals of motion (or conserved quantum numbers,
in quantum theory). This approach is applied to describe
particle creation in
the models of the early universe where the Hubble parameter goes to infinity.
}

\end{abstract}
\newpage

\section{Introduction}

All modern relativistic field theories describe constrained  systems
with a set of "superfluous" degrees of freedom and pure gauge fields.
An identification of physical variables and quantities, in gauge theories and
General Relativity (GR), is a
long-time problem which stimulated Dirac to formulate, for this aim, the
general Hamiltonian theory for constrained systems \cite{d} developed
later by many authors (see e.g. monographs \cite{fs,ht}).
The essence of the Dirac approach to constrained systems
is the reduction of the initial extended phase space to separate
the true physical variables from
the parameters of gauge and general coordinate transformations
and to construct a dynamic equivalent unconstrained system in
the reduced phase space in terms of gauge-invariant variables
which are called the "Dirac observables".

However, the application of the Dirac scheme of the Hamiltonian reduction
to field theories invariant with respect to the reparametrization
of time (GR and cosmological models) meets with an additional
peculiarity: in these theories there are superfluous variables
which are excluded by constraints and abandon the sector of physical
variables (i.e. "Dirac observables"),
but not the sector of measurable quantities
\footnote{It is worth to recall that yet in 1873 J.C.Maxwell wrote \cite{mx}
"The most important aspect of any phenomenon from mathematical point of view
is that of a measurable quantity. I shall therefore
consider electrical phenomena chiefly with a view to their measurement
describing the methods of measurement and defining the standards
on which they depend."} \cite{KPP,grg,plb}.

Examples of such variables are the cosmic scale factor and the invariant
proper time treated as measurable quantities in cosmological models.
In this case, set of equations of the reduced theory
does not reflect all the physical content of the initial extended theory and
must be supplemented by two equations
for the superfluous variable and its momentum \cite{grg}. The second
equation leads to the Friedmann-Hubble law of evolution of the universe
(i.e. the dependence of the red shift of spectral lines
of the cosmic object atoms on a distance of this object from the Earth);
while the first one determines the evolution of measurable density of matter
\cite{plb}.

In the present paper, we would like to reconsider the problem of "particles"
and "quasiparticles"  in the early universe \cite{lp,z,gs,gm}
in the context of the Dirac approach to constrained systems
supplemented by the sector of measurable quantities \cite{KPP,grg,plb}.

The basic idea is the definition of "particles" as
field variables in the holomorphic representation \cite{fs}
which diagonalize the measurable density in the supplemented equations
of the evolution of the universe.

To get the set of integrals of motion (or conserved quantum numbers
in quantum theory), we apply the Bogoliubov transformations
of particle variables for diagonalization of the equations of motion
(but not only of the initial Hamiltonian \cite{lp,z,gs,gm}),
in the correspondence with the original
idea in Bogoliubov's paper \cite{b} where these transformations
were considered as an effective mathematical tool to construct
the energy spectrum of a weakly non-ideal Bose gas.

The Bogoliubov quasiparticles are determined as field variables
with conserved "numbers of quasiparticles".

In the next section, we formulate the problem  and the model.
In Section 3, measurable particles are defined.
In Section 4, we introduce the Bogoliubov transformations to
diagonalize equations of motion.
Section 5 is devoted to the description of creation of
massive particles and gravitons in the early universe.

\section{Statement of  Problem}

General Relativity
\be
\label{gr}
W^{gr}(g)= \int d^4x[-\sqrt{-g}\frac{\mu^2}{6} R(g)+{\cal L}_{\mbox matter} ] ~~~
(\mu^2=M^2_{Planck}\frac{3}{8\pi} )
\ee
is considered as a constrained system for the Dirac-ADM $3+1$ foliation
of the four-dimensional manifold \cite{ADM}
\be
\label{dse}
  (ds)^2_e=g_{\mu\nu}dx^\mu dx^\nu= N^2 dt^2-{}^{(3)}g_{ij}\breve{dx}{}^i
  \breve{dx}{}^j\;;\;\;(\breve{dx}{}^i=dx^i+N^idt).
\ee
The matter Lagrangian in the action describes a set of boson and fermion
fields.
The problems of initial data and
classical and quantum Hamiltonian dynamics of these fields in GR
are conventionally formulated in terms of the so-called Lichnerowicz
conformal-invariant field variables and metric
\cite{L,Y}
\be \label{LYK}
{}^{(n)}f_c(x)= {}^{(n)} f(x) ||{}^{(3)}g(t,x)||^{-n/6};~~~
(ds)_c^2=N_c^2 dt^2-{}^{(3)}g_{(c)ij}\breve{dx}{}^i \breve{dx}{}^j;~~
||{}^{(3)}g_c||=1,
\ee
where $(n= -1, -3/2, 0, 2)$ is the conformal weight for
scalar, vector, spinor, and tensor fields, respectively,
$N_{c}$ is the conformal lapse function, $ds_c$ is the conformal interval.

As a result, the action of GR in terms of the conformal
variables (\ref{LYK}) has the structure \cite{grg} of the
Penrose-Chernikov-Tagirov action \cite{pct} for the conformal scalar field
\be \label{grtot}
W^{gr}_{tot}(\vh_g,f_c)=
\int dt d^3x[-N_c\frac{\vh_g^2}{6}{}^{(4)}R_c+
\vh_g \partial_\mu(N_c\partial^\mu\vh_g)+
N_c({\cal L}_{(\vh_g=0)}-\vh_{g}F+\vh_{g}^2B)],
\ee
where ${\cal L}_{(\vh_g=0)}$ is the massless part of the matter Lagrangian,
$B$ and $F$ are the mass terms of the boson and fermion
fields, respectively,  both the
Newton constant and masses of elementary particles are formed by
the metric scale field $\vh_g$
\be \label{vh1}
\vh_g=\mu ||{}^{(3)}g||^{1/6}.
\ee
Action (\ref{grtot}) can be considered also as scalar
version \cite{grg,plb} of the Weyl conformal theory \cite{HW}, where
the metric scale field (\ref{vh1}) plays the role of the Lichnerowicz
variable (\ref{LYK}) of a scalar field considered as the measure of a change
of the length of a vector in its parallel transport and, simultaneously,
as a modulus of the Higgs scalar field in the Standard Model of
unification of electroweak and strong interactions.
Dynamics of both the theories (the Einstein GR and the Weyl one) is the same,
but not standards of measurement. An Einstein observer measures the
absolute lengths $(ds)_e$, while a Weyl observer  can measure only
the ratio of lengths of two vectors $(ds)_w=(ds_1)_e/(ds_2)_e=(ds_1)_c/(ds_2)_c$
which is conformal-invariant.

The Dirac-ADM-action and
interval (\ref{dse})
are invariant with respect to transformations of a kinemetric subgroup of
the group of general coordinate transformations
\cite{vlad,grg,ps}
\be \label{gt}
t \rightarrow  t'=t'(t);~~~~~
x_{i} \rightarrow  x_{i}'=x_{t}'(t,x_{1},x_{2},x_{3}),
\ee
This group of diffeomorphism of the Hamiltonian
description of GR
includes reparametrizations of the coordinate time which
could not be treated as the "observable" evolution parameter of the
reparametrization-invariant Hamiltonian dynamics of the universe.

One of the main problems of the Hamiltonian reduction
in GR is to pick out the global variable
which can represent the internal
evolution parameter of the corresponding reduced
system.

There is a lot of speculations on this subject,
here we try to continue these attempts in the spirit of
identification of this "internal evolution parameter"
in GR with the  global component of the metric scale field (\ref{vh1})
(or, the scalar field in the Weyl theory \cite{plb})
\cite{Y,grg,plb}
\be \label{vh}
\vh_g(t,x)= \mu ||{}^{(3)}g(t,x)||^{1/6}=\vh_0(T) a(T,x).
\ee
Here $T$ is the world conformal time formed by the global component
$N_0$ of the conformal lapse function:
\be \label{tgr}
N_c(t,x)=N_0(t){\cal N}(T,x);~~~~~~~dT=N_0(t)dt.
\ee
Fields ${\cal N}(T,x);~~a(T,x)$ begin from unit, ${\cal N}(T,x) =1+...$,
in the cosmological perturbation theory \cite{kodama} and form
interactions of the fields with the positive contribution to the
energy constraint.

When interactions are neglected ($ a={\cal N}=1;~~N^k=0;~
g^c_{ij}=\delta_{ij}+h_{ij} $),
the GR action is reduced to the action of the well-known system
of "free" conformal fields in a finite space-volume $(V=\int d^3x)$
\cite{lp,z,gs,gm}
\be \label{e0}
W^E[P_f, f;P_0, \vh_0|t]= \int\limits_{t_1}^{t_2}dt
\left( \left[\int d^3x \sum\limits_{f}P_f\dot f \right]
-\dot \vh_0P_0-N_{0}[-\frac{P_0^2}{4V}+H_0]
+\frac{1}{2}\ik_0(P_0\vh_0)\right),
\ee
where $H_0$ is a sum of the Hamiltonians of "free" fields,
graviton (h), photon (p), massive vector (v), and spinor (s)~)
\be \label{H0}
H_0=H_M+H_R+H_h,
\ee
\be \label{hm}
H_M= \frac{1}{2} \int d^3x\left(P_{(v)}^2+(\ik_iv)^2 +(y_b\vh_0)^2 v^2\right)+
\int d^3x \bar \psi(y_s\vh_0-i\gamma_j\ik_j)\psi,
\ee
is the Hamiltonian of massive conformal fields,
where the role of masses is played by the
"internal evolution parameter" $\vh_0$ multiplied by  dimensionless
constants $y$, $H_R$ is the Hamiltonian
of massless fields with $y=0$, and $H_h$ is the Hamiltonian of
gravitons
\be \label{ho}
H_h=\int d^3x\left(\frac{6(P^T_{(h)})^2}{\vh_0^2}+\frac{\vh_0^2}{24}
(\ik_ih^T)^2\right);~~(h^T_{ii}=0;~~\ik_jh^T_{ji}=0),
\ee
(where the last two equations follow from the separation of the determinant
of three-dimensional metric (\ref{LYK}) and from the momentum constraint).
There is a direct correspondence  between a particle in Special Relativity
(SR)
\be \label{e}
W^E=\int d\tau [- p_{\mu}\dot x^{\mu} - \frac{N}{2m}(-p_{\mu}^2+m^2) ];~~~
\mu=(0,1,2,3)
\ee
and the universe in GR: for
the unobservable coordinate times ~~$(\tau \rightarrow t)$;~~
the internal evolution parameter~~$(x_0 \rightarrow \vh_0)$;~~
the reduced phase space variables~~$(x_i,p_i \rightarrow  P_f,f)$;~~
and the proper time ~~$(dT=Nd\tau \rightarrow dT=N_0dt)$.
Recall that the conformal world time $T$  is considered in the Weyl theory
as the time measured by an observer in the comoving frame, while
in the Einstein GR, an observer measures in the comoving frame
the Friedmann world time $(T_f)$ connected with $T$ by the relation
\be \label{cf}
    \mu dT_f(T)=\vh_0(T)dT.
\ee
Finally, we get the following statement of the problems
considered in the present paper:\\
{\bf to fulfil the
Hamiltonian reduction in the theory described by  action (\ref{e0}),
and to quantize the reduced version of the theory.}

\section{ Evolution of the universe and "particles"}

The Hamiltonian reduction of the system described by action (\ref{e0})
means explicit solving of the constraint
\be \label{ech}
\frac{\delta W^E_0}{\delta N_0}=0\,
\Rightarrow\,
H=[-\frac{P_0^2}{4V}+H_0]=0
\Rightarrow\,
(P_0)_{\pm}=\pm 2 \sqrt{VH_0}\equiv\pm H^R.
\ee
This equation has
two solutions which correspond to two reduced systems with the actions
\be
\label{d7}
W^R_{\pm}(P_f, f|
\vh_0)=\int\limits_{\vh_1=\vh_0(t_1)}^{\vh_2=\vh_0(t_2)}
d\vh_0 \left\{\left(\int
d^3x\sum\limits_{f=h^T,p,v,s} P_f\ik_{\vh_o}f\right)\mp H^R\pm
\frac{1}{2}\ik_{\vh_o} (H^R\vh_0)\right\}
\ee
where the scale factor $\vh_0$ plays the role of the evolution parameter
and $H^R$ is the corresponding Hamiltonian of evolution.

The local equations of motion of  systems (\ref{d7}) reproduce
the invariant sector of the initial extended system and determine
the evolution of fields $(P_f,f)$ with respect to the internal
evolution parameter $\vh_0$
$$
P_f(x, t), f(x, t)\,
\rightarrow\,P_f(x, \vh_0), f(x,\vh_0).
$$
The reduced quantum system is described by the Schr\"odinger type
equation
\be \label{srd}
\frac{d}{id\vh_0}\Psi_R(\vh_0|f)=\pm H^R \Psi_R(\vh_0|f)
\ee
which is a square root of the Wheeler-DeWitt equation
$\hat H\Psi_{wdw}=0$.

We should emphasize that no one reduced scheme (the classical one and two
quantum) contains
the "evolution of the universe" in terms of the world time $T$ measured
by an observer in the comoving frame,
as no one reduced scheme contains this time $dT=N_0dt$.

The "evolution of the universe" is the dependence of the world time $T$
on the internal evolution parameter $\vh_0$. This dependence
is described by the equation for
the "superfluous" momentum $P_0$ (which is omitted by the reduced action $W^R$)
\be
\label{d56}
\frac{\delta W^E}{\delta P_0}=0\,
\Rightarrow\,
\left(\frac{d\vh_0}{N_0dt}\right)_{\pm}
=\frac{(P_0)_{\pm}}{2V}
=\pm\sqrt{\rho({\vh_0})};~~~\rho=\frac{H_0}{V};
\ee
A solution of the latter
\be
\label{70}
T({\vh_0})=\int\limits_0^{\vh_0}d\vh
{\rho}^{-1/2}(\vh),
\ee
is treated in cosmology as the theoretical description of the Hubble red
shift of spectral lines of atoms on cosmic objects at
a distance $D$  from the Earth observer:
\be \label{z}
Z(D)=\frac{\vh_0(T_f)}{\vh_0(T_f-D/c)}-1= {\cal H}^f_{Hub}(T_f) D/c+...;
~~~{\cal H}^f_{Hub}(T_f)=\frac{d\vh_0(T_f)}{\vh_0(T_f)dT_f}
\ee
(where ${\cal H}^f_{Hub}(T_f)$ is the Hubble parameter).

The redshift-data $Z(D)$ determine the observational
dependence of the world proper time on the internal evolution
parameter (here, the cosmic scale factor $\vh_0$).

For a relativistic particle described by action (\ref{e})
the dependence of the proper time $T$
(measured by an observer in the comoving frame)
with respect to the internal evolution parameter $(x_0)$
(measured by another observer in the rest frame) is nothing but
the dynamic version of the Lorentz transformation.

There is an essential physical difference between
 Special Relativity (SR) and General Relativity (GR). In SR,
two times (the proper time $T$ and internal evolution
parameter $x_0$) are measured in two different frames
(the comoving and rest ones)
which exclude each other.
While in GR, both the times, $T$ and $\vh_0$, are
measured by an observer in the comoving frame. The relation
between these two times forms
the evolution of the universe (\ref{70}). This evolution depends
on  standards of measurement of the time.

The same GR dynamics (with the same
Wheeler-DeWitt wave function)
corresponds to different cosmological pictures for different
observers:
an Einstein observer, who supposes that he measures an absolute interval,
obtains the Friedmann-Robertson-Walker (FRW) cosmology where the
red shift is treated as expansion of the universe;
a Weyl observer, who supposes that he measures a relative interval $D_c$,
obtains the Hoyle-Narlikar cosmology \cite{N}.
The red shift and the Hubble law in  the Hoyle-Narlikar cosmology \cite{N}
\be
\label{73}
Z(D_c)=\frac{{\vh_0}(T)}{{\vh_0}(T-D_c/c)}-1\simeq
{\cal H}^c_{Hub}D_c/c;~~~~~~~~~~\,{\cal H}^c_{Hub}
=\frac{1}{{\vh_0}(T)}\frac{d\vh_0(T)}{dT}=
\frac{\sqrt{\rho(T)}}{\vh_0(T)}
\ee
reflect the change of the size of atoms in the process of evolution
of masses ~\cite{N,KPP}.

Equation (\ref{73})
give the relation
between the present-day value of the scalar field and
the cosmological observations  (the density of matter and the Hubble parameter)
\be
\label{d12a}
\vh_0(T)=\frac{\sqrt{\rho_0(T)}}{{\cal H}^c_{Hub}(T)}.
\ee
This value coincides with the Newton constant (or the Planck mass)
in  limits of the observational data
\be \label{exp}
\frac{\vh_0^2(T=T_0)}{6}=\frac{\mu^2 \Omega_0}{6}=
\frac{M_{Pl}^2\Omega_0}{16\pi}\Rightarrow \vh_0(T_0)=\mu\Omega_0^{1/2}.
\ee
The present-day mean matter density
\be\label{d30}
\rho_b=\Omega_0\rho_{cr};~~~~~~~~~~~~~(\rho_{cr}=
\frac{3{\cal H}^c_{Hub}}{{8\pi}}M_{Pl}^2  )
\ee
is estimated from observational data on
luminous matter ($\Omega_0=0.01$), the flat rotation curves of spiral
galaxies ($\Omega_0=0.1$), and other data~\cite{rpp} ($0.1<\Omega_0<2$).

The comparison of
direct observations of masses and numbers of particles in the universe
with  the density of matter measured by the Hubble law (\ref{z}), or
(\ref{73}), in the context of quantum field theory,
means the particle-like treatment of the corresponding Hamiltonian
for free fields $H_0$ which forms this density.
The particle-like representation of $H_0$ has the form
\be \label{hh}
   H_0 = \sum\limits_{k;f=h,p,v,s}\omega_{a_f}(k;\vh_0) \hat {\cal N}_{a_f}
\ee
where $\hat {\cal N}_{a_f}$ is
the "number of particles"
\be \label{part}
\hat {\cal N}_{a_f}=\{a_f^+a_f\}_{\pm}=\frac{1}{2}(a_f^+a_f\pm a_f a^+_f)).
\ee
($(+)$ corresponds to bosons; and $(-)$, to fermions);
and $\omega_{a_f}(k;\vh_0)$ is the one-particle energy
\be \label{om}
\omega_{a_f}(k;\vh_0)=\sqrt{k^2+(y_f\vh_0)^2}
\ee
for gravitons (h) and photons (p) with $(y=0)$, and  for
 bosons (v)  and fermions  (s) with $(y \not =0)$.

Thus, the concepts of  measurable times lead to concepts of
the measurable Hamiltonian and "particles".
In contrast with the conventional approach \cite{lp,z,gs,gm}
(where "particles" are defined as  constant coefficients of
the decomposition of field variables over classical solutions),
we define the "observable particles" as field dynamic variables
in the holomorphic representation which diagonalize
the observable Hamiltonian $H_0$ in the form (\ref{hh}).

This holomorphic representation of field variables \cite{fs} is
\be \label{hp}
f=\sum\limits_{(k,\alpha)}\frac{{\rm exp}(ikx)}{V_0^{3/2}
\sqrt{2\omega_{a_f}(k,\vh_0)}}
(a_f^+(-k,\alpha|t)\epsilon(-k,\alpha)+a_f(k,\alpha|t)\epsilon(k,\alpha)),
\ee
$$
P_f=i\sum\limits_{(k,\alpha)}\frac{{\rm exp}(ikx)
\sqrt{\omega_{a_f}(k,\vh_0)}}{V_0^{3/2}\sqrt{2}}
(a_f^+(-k,\alpha|t)\epsilon(-k,\alpha)-a_f(k,\alpha|t)
\epsilon(k,\alpha))
$$
with operators of creation - $a^+$ and annihilation - $a$ of particles
with the momentum $k$, polarization $\epsilon(k,\alpha)$, and energy
$\omega_{a_f}(k,\vh_0)$,
and $f_h=h^T\sqrt{12\vh_0};~~~P_h={P^T_h}/{\sqrt{12\vh_0}}$.

In terms of the "particle" variables  the extended action (\ref{e0})
has the form
\be \label{ba}
W^E_0=\int\limits_{t_1}^{t_2}dt[-P_0\dot\vh_0+N_0\frac{P_0^2}{4V}-
\sum\limits_{f,k\alpha} (\frac{i}{2}
\bar\chi{}_{a_f}(k,t)\partial_t\chi_{a_f}(k,t)-
N_0\frac{1}{2}\bar\chi{}_{a_f}(k,t){\hat H}_{a_f}(k,t)\chi_{a_f}(k,t))],
\ee
here
\be
\bar\chi{}_{a_f}=(a_f,-a_f^+);\;\;\chi_{a_f}=\left(
\begin{array}{c} a_f^+\\a_f \end{array}\right);\;\;{\hat H}_{a_f}=\left|
\begin{array}{ccc}\omega_{a_f} &,&-i\Delta_f\\ \\ -i\Delta_f&,&-\omega_{a_f}\end{array}
\right|.
\ee
This action includes additional nondiagonal terms describing the creation of
matter in the evolution of the universe:
\be \label{nond}
\begin{array}{|c|} \hline \\
\Delta_p=0;~~~
\Delta_v=\frac{d}{dT}\log \sqrt{\frac{\omega_v(k,T)}{\omega_v(k,0)}};~~~
\Delta_s=\frac{d\vh}{dT}<\frac{k_i\gamma_i \gamma_0 y_s}{2\omega_s(k,T)^2}>;~~
\Delta_h=2\frac{d}{dT}\log(\frac{\vh(T)}{\vh(0)})
\\ \\ \hline
\end{array}
\ee
where $< >$ denotes the matrix element between states with definite spins,
and $\omega(k,0), \vh(0)$ is defined by  initial data.
In the following, for simplicity, we consider photons $(\Delta_p=0)$,
massive bosons $(v)$, and gravitons $(h)$.

On the present-day stage, $\Delta_f\sim 0$ and numbers of all particles are
conserved. In this case, the proper time dynamics with
a fixed number of massive particles $(k=0)$ leads to
the Hubble law for the dust stage of classical universe and
the corresponding wave function of the universe
expressed in terms of the proper time has the form \cite{KPP}
$$
\Psi_R(\vh_0(T)|f)=\exp(iM_{dust}T_f)<f|M_{dust}>;~~~ (~\mu dT_f=\vh_0(T)dT~),
$$
where $M_{dust}$ is a sum of all particle masses of the universe,
$<f|M>$ is the product of oscillator wave functions,
$T_f$ is the Friedmann time and $T$  is the conformal one.
As we have seen above, the quantum universe filled in by a fixed number
($10^{88}$) of massless particles (photons) is described by the wave function
$$
\Psi_R(\vh_0(T)|f)=\exp(iE_{radiation}T)<f|E_{radiation}>
$$
and by the Hubble law of the classical universe in the radiation stage
\cite{KPP}.
Thus, the separation of the global observable time and
energies allows us to unify classical and quantum field theory
with the classical and quantum cosmologies and to give the perspective
of the construction of quantum gravity as a quantum version
of the cosmological perturbation theory \cite{kodama}.

In the early universe,  $\Delta\neq 0$ and numbers of particles are not
conserved
$$
\frac{d\hat {\cal N}_f(T)}{dT}\neq 0.
$$

{\bf The problem is to find integrals of motion of
time-reparametrization system (\ref{ba}) and to describe
creation of "particles" in the evolution of the universe.}

\section{Bogoliubov quasiparticles: integrals of motion}

To construct integrals of motion
 in the context of the above-mentioned Hamiltonian reduction,
we  use the Bogoliubov transformations \cite{b}
of "particle" variables
\be \label{17}
b^+=\alpha^*a^++\beta^*a,\;\;\;\;b=\alpha a+\beta a^+;~~~~
|\alpha|^2-|\beta|^2=1,
\ee
which diagonalize the corresponding classical equations expressed in terms
of "particles" $(a^+,a)$, so that the "number of quasiparticles"  is
 conserved
\be
\frac{d{\cal N}_b(t)}{dt}\equiv \frac{d(b^+b)}{dt}=0.
\ee
The classical equations for action (\ref{ba}) can be written as
\be
i\frac{d}{dT}\chi_a=-{\hat H}_a\chi_a.
\ee
After Bogoliubov transformations (\ref{17})
\be
\chi_b=\hat O\chi_a;\;\;\;
\hat O= \left(
\begin{array}{ll} \alpha^*,&\beta^*\\ \\
\beta,&\alpha\end{array}\right);\;\;\; \hat O{}^{-1}= \left(
\begin{array}{rr} \alpha,&-\beta^*\\ \\ -\beta,&\alpha^*\end{array}\right);
\ee
this equation gets the form
\be
i\frac{d}{dT}\chi_b=[-i\hat O{}^{-1}\frac{d}{dT}\hat O -
\hat O{}^{-1}\hat H{}_a\hat O]\chi_b\equiv -\hat H{}_b\chi_b.
\ee
Let us require  $\hat H{}_b$ to be diagonal
\be
\hat H{}_b=\left(
\begin{array}{rr} \omega_b,&0\\ \\ 0,&-\omega_b\end{array}\right).
\ee
This means that $\alpha$ and $\beta$  satisfy the equations
\be
\omega_b=(|\alpha|^2+|\beta|^2)\omega_a-i(\beta^*\alpha-\beta\alpha^*)\Delta-
i(\beta^*\partial_T\beta-\alpha\partial_T\alpha^*),
\ee
\be
0=2\beta\alpha\omega_a-i(\alpha^{2}-\beta^{2})\Delta-
i(\alpha\partial_T\beta-\beta\partial_T\alpha).
\ee
For
\be
\alpha=\cosh(r)e^{i \theta}\;;\;\;\;\beta=i\sinh(r)e^{-i \theta}
\ee
these equations convert into
\be    \label{29}
\begin{array}{|rl|} \hline &\\
\omega_b=&\omega_a \cosh{2r}-\Delta \sinh{2r} \cos{2\theta}-
\cosh{2r}\partial_T\theta\\  & \\
0=&\omega_a \sinh{2r}-\Delta \cosh{2r} \cos{2\theta}-\sinh{2r}
\partial_T\theta\\
&\\ 0=&\Delta\sin{2\theta}+\partial_Tr \\ & \\ \hline
\end{array}
\ee
If we consider these equations for constant $\omega_a, \Delta$,
we get the result of Bogoliubov paper \cite{b}
\be
\theta=0\;;\;\;\cosh{2r}=\frac{\omega_a}{\omega_b}\;\;\;
\omega_b=\sqrt{\omega_a^2-\Delta^2}.
\ee
Finally , the classical equations in terms of "quasiparticles" are of the form
\be   \label{33}
\frac{d}{dT}b^+=i\omega_bb^+;\;\;\;\frac{d}{dT}b=-i\omega_bb,
\ee
with the solution
\be
b^+=\exp{(iQ)}b^+_0\;;\;\;b=\exp{(-iQ)}b_0;~~~~~Q=\int\limits^T dT' \omega_b(T')
\ee
and the conserved number of quasiparticles
\be \label{imb}
\hat {\cal N}_b(t)=\{b^+(t)b(t)\}=\{b_0^+b_0\}
\ee
where  $b_0^+$ and $b_0$ are  initial data.

To close equation (\ref{33}), we recall that the
evolution of the universe
is determined by the density of "observable particles"
\be \label{obs}
\frac{d\vh}{dT}=\sqrt{\rho(\vh)};\;\;\rho(\vh)=\frac{H_0}{V}=
\frac{\sum\limits_f\omega_f(\vh)\{a_f^+a_f\}}{V_0};
\ee

\be           \label{37}
\begin{array}{|c|} \hline \\
\{a^+a\}=\{b^+b\}\cosh{2r}-\frac{i}{2}(b^{+2}-b^2)\sinh{2r}\\ \\
\hline\end{array}
\ee
Equations (\ref{29})-(\ref{37}) represent a complete set of equations of
classical theory in terms of the holomorphic variables.
This theory has a set of integrals of motion (\ref{imb}) which should be
converted into the quantum numbers of the corresponding quantum reduced system
in agreement with the correspondence principle.

\section{Creation of Particles}

As $\vh_0$ left the phase space to convert into the internal evolution
parameter, in quantum theory we can quantize only  the particle sector
\be
[a,a^+]=1\;;\;\;\;[b,b^+]=1.
\ee
In the following, we restrict ourselves to the  universe in the state
of vacuum of quasiparticles (i.e. the squeezed vacuum):
\be
{}_b<0|b^+b|0>_b=0\;;\;\;<0|\{a^+a\}|0>=\frac{1}{2}\cosh{2r}=
\frac{1}{2}{\cal N}_0(\vh_0).
\ee

The quantum equation for the early universe coincides with
the Schr\"odinger one, and the wave function of the state of
"nothing"  is defined as the squeezed vacuum $b_f(k,0)|0>_b=0$
with the scale factor and the particle numbers
given by eqs. (\ref{obs}) and (\ref{37}).

We can rewrite eqs. (29) in terms of the number of particles ${\cal N}_0(\vh)$
\be \label{41}
\begin{array}{l}
\omega_{b_f}=\omega_f {\cal N}_0-\Delta_f [\sqrt{{\cal N}_0^2-1}\cos{2\theta}+
{\cal N}_02\frac{d\theta}{d\tau_f}] \\ \\
\omega_f \sqrt{{\cal N}_0^2-1}=\Delta_f [{\cal N}_0\cos{2\theta}+\sqrt{{\cal N}_0^2-1}~~
2\frac{d\theta}{d\tau_f}]    \\ \\
\sqrt{{\cal N}_0^2-1}\sin{2\theta}=-\frac{d {\cal N}_0}{d\tau_f};~~~
 \Delta_f=\sqrt{\rho}\frac{d\tau_f(\vh_0)}{2d\vh_0}
\end{array}
\ee
with
\be
\tau_h(\vh_0)=2\log\left(\frac{\vh_0(T)}{\vh_0(0)}\right);\;\;
\omega_h=\sqrt{k^2};~~
\tau_v(\vh_0)=
\log\left(\frac{\omega_v(\vh_0)}{\omega_0}\right);\;\;\
\omega_v(\vh_0)=\sqrt{k^2+(y_v\vh_0)^2}
\ee
where $\omega_0, \vh_0(0)$ are the initial data.
>From first two equations (\ref{41}) we get
$$
\omega_{b_f}=\frac{\Delta_f \cos{2\theta}}{\sqrt{{\cal N}_0^2-1}}.
$$
There are two different regimes $\omega_{a_f} \gg \Delta_f, (\Delta_f=0)$
and $\omega_{a_f} \ll \Delta_f$,
for which eqs.(\ref{41}) can be solved exactly.

For $\Delta_f=0$ we got $\theta=0,~~{\cal N}_0=1$, particles are not created.

In the opposite case, for $\Delta\rightarrow \infty$,
the solution to these equations has the functional form
\be
\theta_f=\frac{\pi}{4};
~~~~{\cal N}_0(\vh_0)=
\frac{1}{2}\left(\frac{\omega_0}{\omega_a(\vh_0)}+
\frac{\omega_a(\vh_0)}{\omega_0}\right)
\ee
for massive particles $(f=v)$, and
\be \label{bh}
\theta_f=\frac{\pi}{4};~~~~
{\cal N}_0(\vh_0)=
\frac{1}{2}\left(\frac{\vh_0^2(0)}{\vh_0^2(T)}+
\frac{\vh_0^2(T)}{\vh_0^2(0)}\right)
\ee
for gravitons $(f=h)$.
In both the cases (v,h), this solution has the zero energy of
the Bogoliubov quasiparticles ($\omega_{b_f}=0$).
In this approximation, the density of gravitons
\be
<\rho(\vh)>=\rho_0\frac{{\cal N}_0(\vh)}{2},
\ee
(where $\rho_0$ is the vacuum density) corresponds to the Friedmann cosmological
model with the "rigid" state in the three-dimensional space with a negative
curvature.

The proper time dynamics is determined by the integral
\be
T(\vh_0)=\int\limits_0^{\vh_0} d\vh
\frac{1}{\sqrt{<\rho(\vh)>}}
\ee
which leads to the following redshift-formula  of
the universe evolution
\be \label{b7}
\frac{\vh_0^2(T)}{\vh^2_0(0)}=
\sinh(\frac{2T}{T_0});~~~~
T_0=\vh_0\sqrt{\frac{2}{\rho_0}}.
\ee
We can see that for a Weyl observer
(with the relative standard of length)
there can be a period of the very fast
inflation-like evolution of the cosmic scale factor,
while  an Einstein-Friedmann
observer (with the absolute standard of the length)
sees the linear dependence of the measured proper time on the scale factor.
Thus, ten billion years for an observer with the absolute standard
can convert into the "biblical"  short period of several thousand years for
an observer with the relative standard.

\section{Conclusion}

We have considered the concepts of "particle" and "quasiparticle"
in the field theory
with the time-reparametrization invariance and the energy constraint,
using as an example the description of the  universe in General Relativity.

The peculiarity of constrained systems is the presence of superfluous degrees
of freedom.
Removing these superfluous degrees of freedom from the constrained theory is
the aim of the Hamiltonian reduction  which
has to represent (according to  the
Dirac theory) an equivalent unconstrained system limited by
the "sector of the Dirac observables".

However, in the considered case the "sector of the Dirac observables"
does not include all "measurable quantities".
The superfluous variables, leaving the phase space,
remain in the theory as measurable quantities out of
the "sector of the Dirac observables" of the reduced version.
In particular, to describe in the generalized Hamiltonian theory
the Hubble law of evolution of the universe
(measured in observational cosmology) we are enforced to remain
in the theory two time-like variables:
the world proper time (measured in the comoving frame)
and cosmic scale factor (as the internal evolution parameter).
While for the description of evolution of the Dirac observables
(i.e. the reduced theory) one invariant time parameter
is sufficient.
The evolution of the universe (as the dependence of the proper time
on the internal evolution parameter) goes beyond the scope of
the "equivalent" reduced theory and determines
the measurable density of the matter field.
"Particles"  are the variables which diagonalize this
 measurable density.
"Quasiparticles" are the variables which diagonalize
the equations of motion and determine a set of integrals of motion.
These definitions strongly  differ from the conventional approach
\cite{lp,z,gs,gm} which goes from the conserved "particles" as initial
data (i.e. constant coefficients of the field decomposition over
classical solutions)
to "unconserved quasiparticles" with diagonalization  of
the initial Hamiltonian, but not the equations of motion.

The Bogoliubov  quasiparticles
classify the states of the universe in
both the classical theory and the quantum one. The states with a fixed
number of quasiparticles are squeezed states for particles.
The Bogoliubov  quasiparticles allow us to define the state of
"nothing"  as the squeezed vacuum.

The formulation of classical and quantum versions  of General
Relativity in terms of measurable quantities leads  to unification of
classical and quantum cosmologies with QFT in GR and
reveals the dependence of the classical evolution of the universe
and the wave function of the universe on "methods of measurement and
their standards".

An Einstein observer measuring lengths by the absolute
standard sees the expansion of the universe, while a Weyl observer
(who can measure the ratio of lengths of two vectors)
sees the Hoyle-Narlikar evolution.

We considered the difference between a Weyl observer and
an Einstein observer
using, as an example, the evolution of the early universe.
A Weyl observer (with relative standard of length) sees the squeezed vacuum
inflation of particle masses due to the zero energy excitation
in the spectrum of quasigravitons, while an Einstein observer sees
the linear dependence of the scale factor on the world time for
the same dynamics.

{\bf Acknowledgments}

\medskip

We are happy to acknowledge interesting and critical
discussions with  B.M. Barbashov, P.N. Bogoliubov,
A. Borowiec, A.V. Efremov,
G.A. Gogilidze, V.G. Kadyshevsky,
A.M. Khvedelidze, E.A. Kuraev,
D. Mladenov,
Yu.G. Palii, V.V. Papoyan,  M. Pawlowski, and G.M. Vereshkov.
One of the authors (V.P.) thanks A. Ashtekar,
C. Isham, T. Kibble, J. Lukierski,
L. Lusanna,  and C. Rovelli for useful discussions.

\end{document}